\documentclass[journal]{IEEEtran}
\IEEEoverridecommandlockouts
\usepackage{cite}
\usepackage{amsmath,amssymb,amsfonts}
\usepackage{algorithmic}
\usepackage{graphicx}
\usepackage{textcomp}
\usepackage[nolist]{acronym}
\usepackage[caption=false]{subfig}
\usepackage{url}

\def\BibTeX{{\rm B\kern-.05em{\sc i\kern-.025em b}\kern-.08em
    T\kern-.1667em\lower.7ex\hbox{E}\kern-.125emX}}

\begin{document}

\begin{acronym}[HKHKJHJKHJKJK]
    \acro{5G}{fifth generation}
    \acro{DL}{downlink}
    \acro{HUBO}{higher-order unconstrained binary optimization}
    \acro{MCS}{modulation and coding scheme}
    \acro{MIMO}{multiple-input multiple-output}
    \acro{MU-MIMO}{multi-user MIMO}
    \acro{MU-MISO}{multi-user multiple-input single-output}
    \acro{NOMA}{non-orthogonal multiple access}
    \acro{OFDM}{orthogonal frequency division multiplexing}
    \acro{OFDMA}{orthogonal frequency division multiple access}
    \acro{PCI}{physical cell identifier}
    \acro{QAOA}{quantum approximate optimization algorithm}
    \acro{QUBO}{quadratic unconstrained binary optimization}
    \acro{RIS}{reconfigurable intelligent surfaces}
    \acro{RB}{resource block}
    \acro{SINR}{signal-to-interference-plus-noise ratio}
    \acro{SNR}{signal-to-noise ratio}
    \acro{SU-MIMO}{single-user MIMO}
    \acro{SVD}{singular value decomposition}
    \acro{TTI}{transmission time interval}
    \acro{UE}{user equipment} 
    \acro{ZF}{zero forcing}
\end{acronym}

\title{QUBO Formulations of the Downlink MIMO Scheduling Problem in 5G Base Stations}

\author{Olli Apilo, and Jorma Kilpi}

\maketitle

\begin{abstract}
Quantum computers can potentially solve large-scale combinatorial problems very efficiently when the problems are first converted into the \ac{QUBO} format. Scheduling in \ac{5G} base stations is a practical combinatorial problem that cannot be solved optimally in real-time using classical computing. We formulate the \ac{DL} \ac{MIMO} scheduling at \ac{5G} base stations as \ac{QUBO} and analyze the \ac{QUBO} formulation scalability with respect to the key system parameters. The \ac{SU-MIMO} \ac{QUBO} formulation looks promising because the number of \ac{QUBO} variables grows linearly while the problem search space grows exponentially with increasing number of users. Based on the simulation results, the suboptimal greedy algorithm for the \ac{SU-MIMO} performs well with a high number of users and low bandwidth. A hybrid approach, where either a quantum solver or a suboptimal low-complexity classical algorithm is selected based on the system parameters, seems sensible in practice. This work paves the way for future quantum and quantum-inspired implementations of scheduling in cellular systems.
\end{abstract}

\begin{IEEEkeywords}
Scheduling, 5G, MIMO, quadratic unconstrained binary optimization, quantum computing
\end{IEEEkeywords}

\section{Introduction}
\label{sec_introduction}

\IEEEPARstart{T}{he} task of the scheduler in \ac{OFDMA}-based cellular communication systems is to select which users are simultaneously served in the given time slot, and which set of subcarriers (i.e., \acp{RB}) are selected for each of the users. In \ac{5G}, each user is restricted to having the same \ac{MCS} and the same number of spatially multiplexed data streams for the given time slot \cite{TS38212}, which tightly couples both link adaptation and \ac{MIMO} mode selection to scheduling. \ac{MIMO} techniques can be divided into \ac{SU-MIMO} where independent data streams are transmitted to a single user, \ac{MU-MIMO} where independent data streams are transmitted to multiple users over the same radio resources, and into \ac{MU-MISO} where the users have only a single antenna, which simplifies the precoding design. When a set of users can be separated in the spatial domain, \ac{MU-MIMO} can bring considerable gain in the system sum rate. However, finding the optimal set of users to be scheduled to the same resource is a combinatorial NP-hard problem. Solving it using exhaustive search becomes computationally prohibitive even for a small number of users \cite{Castaneda17}. Jointly selecting the optimal \ac{RB} allocation for the users, per-user \ac{MCS}, and the number of data streams per-user has not been possible due to extremely large search space. Instead, several sub-optimal lower complexity algorithms have been proposed for \ac{SU-MIMO} \cite{Femenias17}, \ac{MU-MISO} \cite{Ducoing23}, and \ac{MU-MIMO} \cite{Chen23,Wu24}.

Quantum computers can potentially solve large-scale combinatorial problems very efficiently due to their ability to go through the potential solutions in parallel using superposition and entanglement \cite{Chicano25}. The combinatorial problems are typically first converted into the \acf{QUBO} format which is the most widely applied optimization model in quantum computing \cite{Glover19}. When the optimization problem is presented as \ac{QUBO}, it can be directly applied to quantum annealers given that there is enough physical qubits and connectivity between them. \ac{QUBO} problems can also be solved by \ac{QAOA} using the gate-based quantum computers \cite{Volpe25}. However, many of the practical combinatorial problems are neither unconstrained nor of second order. Thus, the practical QUBO formulation often involves transforming linear constraints into quadratic penalty functions \cite{Glover19} and reducing the order of the problem by quadratization \cite{Anthony17}.

Recently, the \ac{QUBO} approach has been successfully applied to various mobile communications problems such as \ac{PCI} planning \cite{Barillaro23}, channel decoding \cite{Kasi20,Kasi24}, \ac{MIMO} detection \cite{Gulbahar25} and precoding \cite{Winter24}, \ac{NOMA} detection \cite{Yonaga25}, and \ac{RIS} optimization \cite{Lim24}. In addition, NTT DOCOMO has already applied quantum annealing to the problem of finding the group of base stations that minimizes the number of paging signals \cite{NttDocomo24}. Formulating the cellular communication system scheduling as a \ac{QUBO} problem has not been previously presented in the literature. Authors of \cite{HsuZhi2026} formulate a simplified \ac{DL} {MU-MISO} problem, where each user is assigned only a single \ac{RB} decoupling the \ac{MCS} selection from scheduling, as a nonlinear integer programming problem and determine a variational quantum algorithm (VQA) to solve it. However, scalability of VQAs is known to be challenging due to measurement shot noise over-head \cite{ScrivaEtAl2024}, \cite{BarligeaEtAl2025}. A related problem of selecting the best set of users for joint transmission, which is equivalent to \ac{MU-MIMO} in the distributed \ac{MIMO} architecture, has been considered in \cite{Volpe24}. In \cite{Volpe24}, the set of possible users for joint transmission is selected from all users by solving a \ac{QUBO} that maximizes the sum \ac{SNR}, while the final selection of users and their power scaling factors are done by a brute force search.

In this paper, we formulate the \ac{MU-MISO} and \ac{SU-MIMO} \ac{DL} scheduling problems in 5G systems as \ac{QUBO} problems and analyze their complexity in terms of the number of required \ac{QUBO} variables, which typically corresponds directly to the number of logical qubits when applied to quantum computers. In addition, we analyze the complexity-performance trade-off of suboptimal scheduling algorithms, which provides insight when solving the problem optimally using quantum computers is of practical interest. We believe that our work will pave the way for future quantum or quantum-inspired implementations of cellular communication system scheduling.

\section{DL scheduling problem in 5G}

\begin{figure}[htbp]
\centerline{\includegraphics[width=\columnwidth]{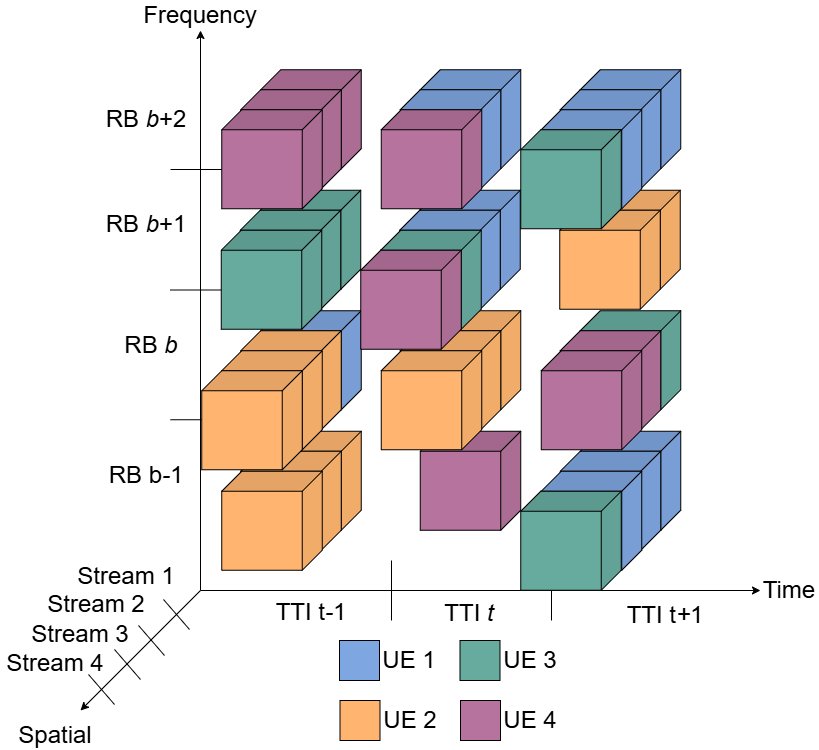}}
\caption{Example of \ac{MU-MIMO} scheduling when $N_T = N_R = K = 4$.} 
\label{fig_schedExample}
\end{figure}

The task of the 5G \ac{DL} scheduler is to allocate the available \acp{RB} for the $K$ \acp{UE} it is serving at each \ac{TTI}. \ac{MIMO} provides additional degrees of freedom for scheduling as the base station has to decide which \acp{UE} and how many data streams per \ac{UE} are multiplexed into the same \ac{RB}. Each \ac{RB} can transmit up to $N_T$ temporally and spatially parallel data streams, where $N_T$ is the number of transmitters at the base station. In Fig.~\ref{fig_schedExample}, where example scheduling decisions are presented when $N_T = 4$, there are at most 4 colored boxes in total in the spatial dimension. For a fixed time slot and frequency, the  colored boxes constitute an \ac{RB}, where the undrawn 'invisible' boxes mean unused capacity in that \ac{RB}. Not every \ac{RB} needs to be fully allocated, nor allocated at all.

The number of antennas at the \ac{UE}, denoted as $N_R$, limits how many parallel data streams can be scheduled for the \ac{UE} in a single \ac{RB}. The number of streams per \ac{UE} has to be fixed for the given \ac{TTI}, i.e. it is not possible to receive different number of streams at different \acp{RB} during the same \ac{TTI}. This is illustrated in Fig. \ref{fig_schedExample}, where the number of spatial streams for the given \ac{UE} and \ac{TTI} is the same over all scheduled \acp{RB}.

In addition, the scheduler selects an \ac{MCS} index for each scheduled \ac{UE}, which also has to be the same for all scheduled \acp{RB} of the given \ac{UE} at the given \ac{TTI}. The \ac{MCS} index, which is a function of received \ac{SINR}, effectively defines how many data bits can be transmitted over an \ac{RB}. The received \ac{SINR} at a \ac{UE} varies depending on the \ac{RB} and on which other \acp{UE} are multiplexed to the same \ac{RB}. This tightly couples conventional scheduling and \ac{MCS} selection.

\subsection{System model}
\label{sec_system}

We consider \ac{DL} scheduling of a single 5G cell where the base station with $N_T$ transmitters is serving $K$ \acp{UE} each having $N_R$ antennas. The available bandwidth is divided into $B$ \acp{RB} which consist of 12 subcarriers. The scheduling decision is done for each \ac{TTI} consisting of 14 \ac{OFDM} symbols. We assume \ac{ZF} precoding for canceling both intra- and inter-user interference from the multiplexed \acp{UE}. To enable simple spatial separation of the transmitted streams, each \ac{UE} estimates its \ac{DL} channel matrix, calculates its \ac{SVD}, and feeds back the equivalent channel matrix to the base station. The \ac{SVD} of $k$th \ac{UE}'s channel matrix $\mathbf{H}_{k,c,m} \in \mathbb{C}^{N_R \times N_T}$ at subcarrier $c$ and \ac{OFDM} symbol $m$ is given by
\begin{equation}
    \mathbf{H}_{k,c,m} = \mathbf{U}_{k,c,m} \mathbf{\Sigma}_{k,c,m} \mathbf{V}_{k,c,m}^H
\end{equation}
where $\mathbf{U}_{k,c,m} \in \mathbb{C}^{N_R \times N_R}$ and $\mathbf{V}_{k,c,m} \in \mathbb{C}^{N_T \times N_T}$ are unitary matrices and $\mathbf{\Sigma}_{k,c,m} \in \mathbb{R}^{N_R \times N_T}$ is a diagonal matrix of singular values. We assume that each \ac{UE} multiplies its received signal by combining matrix $\mathbf{U}_{k,c,m} ^H$. The post-combining equivalent channel matrix is then given as
\begin{equation}
    \label{eq_effectiveChannel}
    \tilde{\mathbf{H}}_{k,c,m}  = \mathbf{U}_{k,c,m} ^H \mathbf{H}_{k,c,m} = \mathbf{\Sigma}_{k,c,m} \mathbf{V}_{k,c,m}^H.
\end{equation}
When $\mathbf{H}_{k,c,m}$ has full rank, $\mathbf{\Sigma}_{k,c,m}$ has $N_R$ nonzero entries, and it is possible to receive up to $N_R$ parallel data streams at each \ac{UE}.

The \ac{SVD} of each \ac{UE}'s channel matrix effectively converts the \ac{MU-MIMO} system into a \ac{MU-MISO} system with $KN_R$ virtual single-antenna users \cite{Panajotovic15}. For the rest of the paper, we use virtual user indexing that maps the $n$th data stream of the $k$th \ac{UE} into virtual user $v=v_{k,n}$ such that 
\begin{equation}
    v_{k,n} = (k-1)N_R + n,
\end{equation}
with $ 1 \leq k \leq K$, and $ 1 \leq n \leq N_R$. A box in Fig.~\ref{fig_schedExample} represents a virtual user. The received signal of $v$th virtual user corresponding to the $n$th stream of \ac{UE} $k$ is then
\begin{equation}
    \tilde{y}_v = \tilde{\mathbf{h}}_v \mathbf{x} + \tilde{n}_v = \sigma_{k,n} \mathbf{v}^H_{k,n} \mathbf{x} + \tilde{n}_v
\end{equation}
where $\tilde{\mathbf{h}}_v \in \mathbb{C}^{1 \times N_T}$ is the equivalent channel vector for virtual user $v$, $\mathbf{x} \in \mathbb{C}^{N_T \times 1}$ is the vector of transmitted precoded symbols, $\tilde{n}_v \sim \mathcal{CN}(0,N_0)$ is the effective additive noise term, $\sigma_{k,n}$ is the $n$th singular value of the channel matrix for \ac{UE} $k$, and $\mathbf{v}_{k,n}$ is the $n$th right singular vector of \ac{UE} k. The subcarrier and \ac{OFDM} symbol indexing have been left out for brevity.

Let us assume the scheduler has selected a set of virtual users $\mathcal{V}$, with 
\begin{equation}
2\leq |\mathcal{V}| \leq \min(N_T,KN_R) 
\end{equation}
to be multiplexed to \ac{RB} $b$. In Fig.~\ref{fig_schedExample}, the boxes of the same color in a fixed RB can be interpreted as being virtual users associated to the same \ac{UE}; the constraint that only $N_R$ virtual users of the same UE can be included in $\mathcal{V}$ still holds. Similar to \cite{Chen23}, we assume constant fading within an \ac{RB}. The combined effective channel matrix for those virtual users in the selected set $\mathcal{V} = \{v_1, \ldots, v_{|\mathcal{V}|}\}$ can be given as
\begin{equation}
    \tilde{\mathbf{H}}_{\mathcal{V},b} = \left[ \tilde{\mathbf{h}}_{v_1,b}^T \cdots \tilde{\mathbf{h}}_{v_{|\mathcal{V}|},b}^T \right]^T \in \mathbb{C}^{|\mathcal{V}| \times N_T}.
\end{equation}
The transmitted symbol vector is then 
\begin{equation}
    \mathbf{x} = \mathbf{F}_{\mathcal{V},b} \mathbf{s}_{\mathcal{V},b}
\end{equation}
where $\mathbf{F}_{\mathcal{V},b} \in \mathbb{C}^{N_T \times |\mathcal{V}|}$ is the precoding matrix with transmit power constraint $\Vert \mathbf{F}_{\mathcal{V},b} \Vert^2_F \leq P_T$ and $\mathbf{s}_{\mathcal{V},b} \in \mathbb{C}^{|\mathcal{V}| \times 1}$ is the vector of data symbols with $E\{|s_{v}|^2\} = 1$. \ac{ZF} precoding over the equivalent channel matrix with equal power allocation results in equal post-combining \ac{SNR} for all the \acp{UE} multiplexed into the same \ac{RB} \cite{Panajotovic15}. The post-combining \ac{SNR} for each virtual user in $\mathcal{V}$ at \ac{RB} b can be given as
\begin{equation}
    \label{eq_snr}
    \gamma_{b}(\mathcal{V}) = \frac{P_T}{N_0 \left\Vert \tilde{\mathbf{H}}_{\mathcal{V},b}^H \left( \tilde{\mathbf{H}}_{\mathcal{V},b} \tilde{\mathbf{H}}_{\mathcal{V},b}^H \right)^{-1} \right\Vert^2_F}.
\end{equation}

The scheduler also selects an \ac{MCS} index out of $M$ possible indices for a \ac{UE} based on the \ac{SNR} values at the \acp{RB} the \ac{UE} has been scheduled. Each \ac{MCS} index corresponds to a certain number of data bits per \ac{RB} $r_m$ \cite{TS38214}. Given the \ac{MCS} index $m_k$ for \ac{UE} k, the number of data bits at \ac{RB} $b$ for the virtual user $v = (k-1)N_R + n, v \in \mathcal{V}$ is given by
\begin{equation}
    \label{eq_rate}
    r_{v,b,m_k}(\mathcal{V}) = 
    \begin{cases}
        r_{m_k}, & \gamma_{b}(\mathcal{V}) \geq \theta_{m_k}\\
        0, & \text{otherwise}
    \end{cases}
\end{equation}
where $\theta_{m_k}$ is the minimum \ac{SNR} at which the \ac{UE} can still successfully decode the data bits for \ac{MCS} index $m_k$. To simplify the analysis, the retransmissions are not considered in this study.

\subsection{Problem definition}

The scheduling problem can be formulated as binary optimization by introducing binary decision variables, $\mathbf{X}$ and $\mathbf{Z}$. $\mathbf{X}$ is a $K N_R \times B$ binary matrix indicating $x_{v,b} = 1$ when virtual user $v$ is scheduled at \ac{RB} $b$. Similarly, $\mathbf{Z}$ is a $M \times K$ binary matrix indicating $z_{m,k} = 1$ when \ac{MCS} index $m$ is selected for \ac{UE} $k$. The \ac{MU-MIMO} scheduling problem can be formulated as
\begin{subequations}
\begin{align}
\max_{\mathbf{X},\mathbf{Z}} \quad & \sum_{b=1}^B \sum_{v=1}^{K N_R} \sum_{m=1}^M \frac{x_{v,b} z_{m,\lceil{v/N_R}\rceil} r_{v,b,m}(\mathbf{x}_b)}{\tilde{R}_{\lceil{v/N_R}\rceil}}
\label{MUMIMO}\\
\textrm{s.t.} \quad & \sum_{v=1}^{K N_R} x_{v,b} \leq N_T, \quad \forall b = 1,\ldots,B \label{constraint1}\\
& \sum_{m=1}^M z_{m,k} \leq 1, \quad \forall k = 1,\ldots,K \label{Constraint3}\\
& \sum_{v=(k-1)N_R+1}^{kN_R} x_{v,b} = \sum_{v=(k-1)N_R+1}^{kN_R} x_{v,c}, \ b \neq c \label{Constraint2a}\\
& \mathrm{OR} \notag\\
& \sum_{v=(k-1)N_R+1}^{kN_R} x_{v,b} = 0, \label{Constraint2b}\\
& \forall k = 1,\ldots,K, \forall b,c = 1,\ldots,B \notag
\end{align}
\end{subequations}
where $\tilde{R}_{\lceil{v/N_R}\rceil}$ is the per-\ac{UE} scaling term for fairness. Constraint (\ref{constraint1}) ensures that the total number of streams does not exceed the number of transmitters at the base station. Constraint (\ref{Constraint3}) is to guarantee that only one \ac{MCS} index is used for a \ac{UE}. Either Constraint (\ref{Constraint2a}) or (\ref{Constraint2b}) has to be fulfilled, which ensures the same number of streams is selected for a \ac{UE} whenever multiple \acp{RB} are scheduled for that \ac{UE}. Note that unlike Constraint (\ref{constraint1}), Constraints (\ref{Constraint3}) and (\ref{Constraint2a}) are not based on physical radio propagation limitations, but they are rather 5G system design choices to keep the \ac{DL} control signaling overhead and \ac{UE} receiver complexity low. For example, assuming that Constraint (\ref{Constraint3}) is removed would result in control signaling overhead that would scale linearly with the number of \acp{RB}. In addition, the \ac{UE} processing complexity would increase due to decoding multiple transport blocks and potentially handling multiple parallel retransmissions per time slot.

\subsubsection{MU-MISO}

In case of \ac{MU-MISO} scheduling, the \acp{UE} have only one antenna and there is one-to-one mapping between the virtual users and \acp{UE}. This simplifies the scheduling problem as $\mathbf{X}$ becomes a $K \times B$ binary matrix and Constraints (\ref{Constraint2a}) and (\ref{Constraint2b}) can be removed.

\subsubsection{SU-MIMO}

In the \ac{SU-MIMO} case the post-combining \ac{SNR} for \ac{UE} $k$ depends only on the number of streams $n \leq N_R$ allocated for that \ac{UE}. If we assume that the singular values in $\mathbf{\Sigma}_k$ are in descending order, the \ac{SU-MIMO} post-combining \ac{SNR} for \ac{UE} $k$ at \ac{RB} $b$ can be given as
\begin{equation}
    \gamma_{k,b}(n) = \frac{P_T \sigma_{n,n}^2}{N_0}.
\end{equation}
The number of data bits per \ac{UE} now scales linearly with the number of data streams:
\begin{equation}
    \label{eq_suMimoRate}
    r_{k,b,m_k}(n) = 
    \begin{cases}
        n r_{m_k}, & \gamma_{k,b}(y_k) \geq \theta_{m_k}\\
        0, & \text{otherwise}
    \end{cases}.
\end{equation}
Let $\mathbf{Y}$ be a $N_R \times K$ matrix indicating $y_{n,k} = 1$ when $n$ data streams are scheduled for \ac{UE} $k$. The \ac{SU-MIMO} scheduling problem can then be given as
\begin{subequations}
\begin{align}
\label{eq_suMiMoProblem}
\max_{\mathbf{X},\mathbf{Y},\mathbf{Z}} \quad & \sum_{b=1}^B \sum_{k=1}^K \sum_{n=1}^{N_R} \sum_{m=1}^M \frac{x_{k,b} y_{n,k} z_{m,k} r_{k,b,m}(n)}{\tilde{R}_k}\\
\textrm{s.t.} \quad & \sum_{k=1}^K x_{k,b} \leq 1, \quad \forall b = 1,\ldots,B \label{constraint1_SUMIMO}\\
& \sum_{m=1}^M z_{m,k} \leq 1, \quad \forall k = 1,\ldots,K \label{Constraint3_SUMIMO}\\
& \sum_{n=1}^{N_R} y_{n,k} \leq 1, \quad \forall k = 1,\ldots,K \label{Constraint2_SUMIMO}
\end{align}
\end{subequations}
where Constraint (\ref{Constraint2_SUMIMO}) guarantees that the same number of data streams is selected for a \ac{UE} for all \acp{RB} scheduled for it.

\subsection{Algorithms}
\label{sec_alg}

In this section, the optimal brute force search algorithms and several sub-optimal variants are discussed for the scheduling problem. The calculation of the \ac{SNR} values is essentially the same for all the algorithms, and its complexity is not included in the analysis.

\subsubsection{MU-MIMO}

Solving the \ac{MU-MIMO} scheduling problem by exhaustive search is an extremely complex problem. The search should go through all combinations of \ac{MCS} and data stream assignments for each \ac{RB}. The search space size is presented as $(M N_R)^K B \left( \sum\limits_{k=1}^{\min(K,N_T)} \binom{K}{k} \right)$ in \cite{Chen23}. However, if the scheduled number of data streams for user $k$ is $n$, there are $\binom{N_R}{n}$ ways to select the scheduled virtual \acp{UE} for that user, and the optimal selection depends on which other users are scheduled on the same \ac{RB}. Thus, the actual search space is larger than that presented in \cite{Chen23}. In practice, once $\mathbf{Y}$ is fixed in the outer loop, the scheduler has to go through all the possible sets $\mathcal{V}_b$ fulfilling Constraints (\ref{Constraint2a}) or (\ref{Constraint2b}).

The search space size would be exactly \linebreak $(M N_R)^K B \left( \sum\limits_{k=1}^{\min(K,N_T)} \binom{K}{k} \right)$ if only the first $n$ virtual \acp{UE} of user $k$ are considered for scheduling when $y_{n,k} = 1$ and $y_{l,k} = 0, \forall l \neq n$. This suboptimal algorithm is expected to perform well when $K >> N_R$. The complexity can be reduced by choosing the \ac{MCS} indices for the users only after the scheduling decision is made. In this sub-optimal algorithm, the best $\mathcal{V}_b$ is first searched for each \ac{RB}. Then for each \ac{UE}, the largest \ac{MCS} index resulting in non-zero data rate for all \acp{RB} scheduled for the given \ac{UE}, is selected. In this case, the search space size reduces to $N_R^K B \left(\sum\limits_{k=1}^{\min(K,N_T)} \binom{K}{k}\right)$.

The complexity of the \ac{MU-MIMO} scheduling problem can be further reduced by the greedy algorithm that schedules an \ac{RB} to the virtual \ac{UE} set with the highest rate at that \ac{RB}. This fixes the number of data streams for the users in the scheduled virtual \ac{UE} set. Any further scheduling decisions must fulfill the data stream requirements given in Constraints (\ref{Constraint2a}) and (\ref{Constraint2b}). Let $\mathcal{W}$ be the set of all possible sets $\mathcal{V}$, i.e. all possible virtual \ac{UE} allocations within \ac{RB}. The number of possible virtual \ac{UE} allocations is 
\begin{equation}
    |\mathcal{W}| = \sum_{v=1}^{\min(KN_R,N_T)} \binom{KN_R}{v} - K \sum_{n=1}^{N_R} \binom{N_R}{n} + KN_R
\end{equation}
where the last two terms ensure that only the \ac{SU-MIMO} allocations with the highest \ac{SNR} are included. The allowed virtual \ac{UE} allocations for the given user $k$ and the given number of data streams $n = 0,1, \ldots, N_R$ are denoted by $\mathcal{W}_{k,n} \subset \mathcal{W}$, $\mathcal{W}_{k,0} \bigcup \mathcal{W}_{k,1} \bigcup \ldots \bigcup \mathcal{W}_{k,N_R} = \mathcal{W}$. The greedy \ac{MU-MIMO} scheduling algorithm is given as
\begin{algorithmic}[1]
    \STATE Solve $\hat{m}_{\mathcal{V},b}$ from $\theta_{\hat{m}_{\mathcal{V},b}} \leq \gamma_{b}(\mathcal{V}) < \theta_{\hat{m}_{\mathcal{V},b}+1}$, $\forall b = 1,\ldots,B, \forall \mathcal{V} \in \mathcal{W}$
    \STATE $\mathbf{X} \gets 0_{KN_R \times B}$, $\mathbf{Z} \gets 0_{M \times K}$, $\mathbf{n} \gets 0_{K \times 1}$
    \STATE $N_{RB} \gets 0$
    \STATE $\tilde{m}_k \gets 29$, $\forall k = 1, \dots, K$    
    \STATE $\mathcal{B} \gets \{1, \ldots, B\}$
    \WHILE{$N_{RB} < B$}
        \STATE $R_{\text{max}} \gets 0$
        \FOR{each $b \in \mathcal{B}$}
            \FOR{each $\mathcal{V} \in \mathcal{W}$}
                \STATE $\hat{R} \gets 0$
                \FOR{each $v \in \mathcal{V}$}
                    \STATE $\hat{R} \gets \hat{R} + r_{b,\hat{m}_{\mathcal{V},b}}(\mathcal{V}) / \tilde{R}_{\lceil{v/N_R}\rceil}$
                \ENDFOR
                \IF{$\hat{R} > R_{\text{max}}$}
                    \STATE $\hat{b} \gets b$, $\hat{\mathcal{V}} \gets \mathcal{V}$
                    \STATE $R_{\text{max}} \gets \hat{R}$
                \ENDIF
            \ENDFOR            
        \ENDFOR
        \STATE $\mathcal{B} \gets \mathcal{B} - \hat{b}$ 
        \STATE $N_{RB} \gets N_{RB} + 1$
        \FOR{each $v \in \hat{\mathcal{V}}$}                        
            \STATE $x_{v,\hat{b}} \gets 1$            
            \STATE $k \gets \lceil{v/N_R}\rceil$
            \IF{$n_k = 0$}
                \STATE $n_k \gets n_k + 1$
            \ENDIF
            \IF{$\hat{m}_{\hat{\mathcal{V}},\hat{b}} < \tilde{m}_k$}
                \STATE $\tilde{m}_k \gets \hat{m}_{\hat{\mathcal{V}},\hat{b}}$
            \ENDIF
        \ENDFOR  
        \FOR{$k = 1, \ldots, K$}
            \IF{$n_k > 0$}
                \STATE $\mathcal{W} \gets \mathcal{W} - \mathcal{W}_{k,i}$, $i > 0$ and $i \neq n_k$ \label{streamConstraint}
            \ENDIF
        \ENDFOR
    \ENDWHILE
    \FOR{$k = 1, \dots, K$}
        \STATE $z_{\tilde{m}_k,k} = 1$
    \ENDFOR
    \RETURN $\mathbf{X}$, $\mathbf{Z}$
\end{algorithmic}
Algorithm line \ref{streamConstraint} effectively reduces the search space whenever a new user is scheduled and guarantees that the data stream constraint is fulfilled for the given user. In the worst case, all the \acp{RB} are scheduled for the same user and all combinations of virtual \ac{UE} allocations for the other $K-1$ users must be considered at each iteration, which results in the worst-case search space size of $B|\mathcal{W}| + \left(|\mathcal{W}| - \sum_{n=1}^{N_R}\binom{N_R}{n}\right) \frac{B(B-1)}{2}$. 

\subsubsection{MU-MISO}

The \ac{MU-MISO} scheduling problem can be solved by going through all possible combinations of \ac{MCS} assignments, and then finding which set $\mathcal{V}_b$ achieves the highest data rate when scheduled for the given \ac{RB} $b$. The complexity of this brute force approach grows exponentially with $K$ as the search space size becomes $M^K B \left(\sum\limits_{k=1}^{\min(K,N_T)} \binom{K}{k}\right)$.

As $M$ can be up to 29 in 5G \cite{TS38214}, a straightforward way to reduce the complexity is to select the per-\ac{UE} \ac{MCS} index only after the scheduling decision is done. In this sub-optimal algorithm, the best $\mathcal{V}_b$ is first searched for each \ac{RB}. Then for each \ac{UE}, the largest \ac{MCS} index resulting in non-zero data rate for all \acp{RB} scheduled for the given \ac{UE}, is selected. The search space size reduces to only $B \left(\sum\limits_{k=1}^{\min(K,N_T)} \binom{K}{k}\right)$.

\subsubsection{SU-MIMO}
\label{sec_alg_sumimo}

Like the \ac{MU-MISO} scheduling problem, the \ac{SU-MIMO} scheduling problem can be solved by going through all possible combinations of per-\ac{UE} data stream allocations and \ac{MCS} assignments, and then finding which \ac{UE} achieves the highest data rate when scheduled for the given \ac{RB}. The complexity of the brute force approach again grows exponentially with $K$ as the search space size is $(N_R M)^K BK$. In practice the search space can be somewhat reduced because only the \ac{MCS} indices in the range $\{m_{\text{min},k}, \ldots, m_{\text{max},k}\}$ should be considered for \ac{UE} $k$ with the given number of data streams. The minimum and maximum relevant \ac{MCS} indices for \ac{UE} $k$ with $n$ data streams fulfill
\begin{equation}
    \theta_{m_{\text{min},k}} \leq \min \{ \gamma_{k,1}(n), \ldots, \gamma_{k,B}(n) \} < \theta_{m_{\text{min},k} + 1}
\end{equation}
and
\begin{equation}
    \label{eq_Mmax}
    \theta_{m_{\text{max},k}} \leq \max \{ \gamma_{k,1}(n), \ldots, \gamma_{k,B}(n) \} < \theta_{m_{\text{max},k} + 1},
\end{equation}
respectively. Like the \ac{MU-MISO} case, the \ac{MCS} indices can be decided only after the scheduling decision is made. This sub-optimal approach results in the search space size of $N_R^KBK$.

The complexity of the \ac{SU-MIMO} scheduling can be further reduced by the greedy approach where an \ac{RB} is scheduled to the \ac{UE} with the highest rate at that \ac{RB}. The first \ac{RB} scheduled for a \ac{UE} decides the number of data streams for that \ac{UE} and fixes the value of $\mathbf{y}_k$. The greedy \ac{SU-MIMO} scheduling algorithm is given below
\begin{algorithmic}[1]
    \STATE Solve $\hat{m}_{k,b,n}$ from $\theta_{\hat{m}_{k,b,n}} \leq \gamma_{k,b}(n) < \theta_{\hat{m}_{k,b,n}+1}$, $\forall k,b,n$
    \STATE $\hat{R}_{k,b,n} \gets r_{k,b,\hat{m}_{k,b,n}}/\tilde{R}_k$, $\forall k,b,n$
    \STATE $\mathbf{X} \gets 0_{K \times B}$, $\mathbf{Y} \gets 0_{N_R \times K}$, $\mathbf{Z} \gets 0_{M \times K}$
    \STATE $N_{RB} \gets 0$
    \STATE $\tilde{m}_k \gets 29$, $\forall k = 1, \dots, K$
    \WHILE{$N_{RB} < B$}
        \STATE $R_{\text{max}} \gets 0$
        \FOR{$b = 1, \dots, B$}
            \FOR{$k = 1, \dots, K$}
                \FOR{$n = 1, \dots, N_R$}
                    \IF{$\hat{R}_{k,b,n} > R_{\text{max}}$}
                        \STATE $\hat{b} \gets b$, $\hat{k} \gets k$, $\hat{n} \gets n$
                        \STATE $R_{\text{max}} \gets \hat{R}_{k,b,n}$
                    \ENDIF
                \ENDFOR
            \ENDFOR
        \ENDFOR
        \STATE $x_{\hat{k},\hat{b}} \gets 1$, $y_{\hat{n},\hat{k}} \gets 1$
        \IF{$\hat{m}_{\hat{k},\hat{b},\hat{n}} < \tilde{m}_k$}
            \STATE $\tilde{m}_k \gets \hat{m}_{\hat{k},\hat{b},\hat{n}}$
        \ENDIF
        \STATE $\hat{R}_{k,\hat{b},n} = 0$, $\forall k,n$ \label{line_rb}
        \STATE $\hat{R}_{\hat{k},b,l} = 0$, $\forall l \neq \hat{n}$, $\forall b$ \label{line_streams}
        \STATE $N_{RB} \gets N_{RB} + 1$
    \ENDWHILE
    \FOR{$k = 1, \dots, K$}
        \STATE $z_{\tilde{m}_k,k} = 1$
    \ENDFOR
    \RETURN $\mathbf{X}$, $\mathbf{Y}$, $\mathbf{Z}$
\end{algorithmic}
Algorithm line \ref{line_rb} ensures that the same \ac{RB} is scheduled to only one \ac{UE}. Line \ref{line_streams} ensures that the same number of data streams is allocated for the scheduled \ac{UE} over all \acp{RB}. When the search is done only over the non-zero elements of $\hat{R}_{k,b,n}$, the search space reduces at each iteration of the while loop. The search space size for the greedy \ac{SU-MIMO} scheduling algorithm is $N_RKB + (N_R(K-1) + 1) \frac{B(B-1)}{2}$ in the worst case. 
The worst case occurs when the first $B-1$ \acp{RB} are all scheduled to the same \ac{UE}.

\section{QUBO formulation of the problem}
\label{sec_qubo}

In resource allocation or scheduling problems formulated as \ac{QUBO} we have binary {\em decision} variables $\xi_i$, $i=1,\ldots,|I|$, one variable for each resource unit $i$ decision. We use the notation $I=\{1,2,3,\ldots,|I|\}$ for the index set with cardinality $|I|$. We consider a mathematical expression of the form
\begin{equation}
    \xi^{T}Q\xi=\sum_{i=1}^{|I|}\sum_{j=1}^{|I|}q_{ij}\xi_i\xi_j,
    \label{quboFormula}
\end{equation}
and we want to find the vector $\xi=(\xi_1,\ldots,\xi_{|I|})$ that minimizes the expression (\ref{quboFormula}). The elements of the matrix $Q=(q_{ij})$ are coefficients that depend on the problem.

Optimization problems usually have constraints. If the constraints can be reformulated as quadratic penalty terms of the binary variables, the constraints are embedded in \ac{QUBO} 
and we are in the 'unconstrained' situation. The constraints can reduce the set of solutions or make it harder to find any.

\subsection{MU-MIMO}
\label{seq_qubo_muMimo}

First, we briefly consider a QUBO approach to the general MU-MIMO problem. Recall that the numbers $N_T$, $N_R$, $B$, and $M$ are parameters. Only $K$, the number of \acp{UE}, is variable. The number of virtual streams is $N_RK$. If $N_RK\leq B$, then each virtual stream could have its own $b$. If $N_RK>N_TB$, then all virtual streams cannot be allocated in the same time slot \ac{TTI}.

The term $r_{v,b,m}(\mathbf{x}_b)$  in (\ref{MUMIMO}) requires a careful approach, since the rate in RB $b$ depends on all users (virtual streams) that are allocated to this $b$, the set $\mathcal{V}_b$, which is a subset of the $b$th column of the $N_RK\times B$ matrix $\mathbf{X}$. This means that, for a fixed resource block $b$, we should study products of binary variables of the form
\begin{equation}
    \prod_{u=1}^{\min\{N_T,N_RK,|\mathcal{V}_b|\}} x_{u,b},
    \label{quboTerm}
\end{equation}
and if $|\mathcal{V}_b|>2$ this leads to problems of higher-order than quadratic. The order of the \ac{MU-MIMO} scheduling problem can be as high as $N_T + 1$ where the additional dimension comes from the \ac{MCS} index decision variable $z_{m,k}$. Even though there are systematic procedures available for converting higher-order problems to quadratic \cite{Anthony17}, the number of additional auxiliary variables soon becomes very high for $N_T > 2$. 

The \ac{MU-MIMO} scheduling problem would be relatively straightforward to formulate as \ac{QUBO} if Constraints (\ref{Constraint3}), (\ref{Constraint2a}), and (\ref{Constraint2b}) could be relaxed. These constraints are per-\ac{UE} over all the \acp{RB} while our approach in the \ac{MU-MIMO} case first converts the system into the \ac{MU-MISO} system with virtual users. This makes the proper formulation of the constraints as quadratic penalties difficult. Because of these challenges and the high-order of the binary optimization problem, we focus in this paper on the simpler \ac{MU-MISO} and \ac{SU-MIMO} cases.

\subsection{MU-MISO}
\label{sec_muMiso}

A straightforward way to formulate the \ac{MU-MISO} scheduling problem as quadratic binary optimization would be to introduce a binary $|\mathcal{W}| \times B$ matrix $\mathbf{A}$ indicating which of the \ac{UE} sets in $\mathcal{W}$ is scheduled for a given \ac{RB}. The number of different \ac{UE} sets in the \ac{MU-MISO} case is 
\begin{equation}
    |\mathcal{W}| = \sum_{v=1}^{\min(K,N_T)} \binom{K}{v}.
\end{equation}
The mapping of the binary decision variables $\mathbf{\xi}$ to $\mathbf{A}$ and $\mathbf{Z}$ would then be
\begin{equation}
    \label{eq_quboVariables}
    \mathbf{\xi}^T = \left[\mathbf{a}_1^T \cdots \mathbf{a}_B^T \quad \mathbf{z}_1^T \cdots \mathbf{z}_K^T \right]^T.
\end{equation}
The elements of the \ac{QUBO} matrix can be expressed with the help of (\ref{eq_rate}) as
\begin{equation}
    q_{ij} =
    \begin{cases}
        -\frac{r_{k,b,m}(\mathcal{V}_w)}{\tilde{R}_k}, &i = 1, \ldots, B|\mathcal{W}|,\\
        &j = B|\mathcal{W}|+1, \ldots, B|\mathcal{W}| + KM,\\
        &k \in \mathcal{V}_w\\
        0, &\text{otherwise}
    \end{cases}
\end{equation}
where $b = \lceil{i/|\mathcal{W}|}\rceil$, $w = i - (b-1)|\mathcal{W}|$,  $k = \lceil{(j - B|\mathcal{W}|)/M}\rceil$, and $m = j - B|\mathcal{W}| - (k-1)M$.

The \ac{MU-MISO} scheduling problem is constrained by the fact the only a single \ac{UE} set can be scheduled for an \ac{RB} and by (\ref{Constraint3}). When rewriting them using the \ac{QUBO} variables we get
\begin{equation}
    \label{eq_constraint1_muMiso}
    \sum_{i = (b-1)|\mathcal{W}| + 1}^{b|\mathcal{W}|} \xi_i \leq 1, \quad \forall b = 1,\ldots,B
\end{equation}
and
\begin{equation}
    \label{eq_constraint3_muMiso}
    \sum_{j = B|\mathcal{W}| + (k-1)M + 1}^{B|\mathcal{W}| + kM} \xi_j \leq 1, \quad \forall k = 1,\ldots,K.    
\end{equation}
Constraints (\ref{eq_constraint1_muMiso}) and (\ref{eq_constraint3_muMiso}) can be converted to quadratic penalty terms by using a simple transformation from \cite{Glover19}. The resulting \ac{QUBO} problem can be given as
\begin{equation}
    \label{eq_muMisoQubo}
    \begin{split}
        \min_{\mathbf{\xi}} \mathbf{\xi}^T \tilde{\mathbf{Q}} \mathbf{\xi} &= \mathbf{\xi}^T \mathbf{Q} \mathbf{\xi} + \sum_{i = (b-1)|\mathcal{W}| + 1}^{b|\mathcal{W}| - 1} \sum_{j = i+1}^{b|\mathcal{W}|} \lambda_1 \xi_i \xi_j + \\
        &\sum_{l = B|\mathcal{W}| + (k-1)M + 1}^{B|\mathcal{W}| + kM - 1} \sum_{o = l+1}^{B|\mathcal{W}| + kM} \lambda_2 \xi_l \xi_o, \\
        & \forall b = 1,\ldots,B, \quad k = 1,\ldots,K
    \end{split}
\end{equation}
where $\tilde{\mathbf{Q}}$ is the final \ac{QUBO} matrix including the penalty terms. The number of \ac{QUBO} variables is $|\mathcal{W}|B + MK$.

\subsection{SU-MIMO}
\label{sec_suMimoQubo}

The \ac{SU-MIMO} scheduling problem (\ref{eq_suMiMoProblem}) has three binary decision variable matrices $\mathbf{X}$, $\mathbf{Y}$, and $\mathbf{Z}$ that can be mapped into \ac{HUBO} variables as
\begin{equation}
    \mathbf{\xi}^T = [\mathbf{x}_1^T \cdots \mathbf{x}_B^T \quad \mathbf{y}_1^T \cdots \mathbf{y}_K^T \quad \mathbf{z}_1^T \cdots \mathbf{z}_K^T]^T.
\end{equation}
The \ac{SU-MIMO} scheduling problem is of the third order with the \ac{HUBO} coefficients from (\ref{eq_suMimoRate})
\begin{equation}
    q_{ijl} =
    \begin{cases}
        -r_{k,b,m}(n)/\tilde{R}_k, &i = 1, \ldots, BK,\\
        &j = BK + 1, \ldots, BK + KN_R,\\
        &l = K(B + N_R) + 1, \ldots,\\
        & \quad \ \  K(B + N_R + M),\\
        & i - (\lceil{i/K}\rceil - 1)K =\\
        &\lceil{(j - BK)/N_R}\rceil =\\
        &\lceil{(l - BK - N_RK)/M}\rceil\\
        0, &\text{otherwise}
    \end{cases}
\end{equation}
where $b = \lceil{i/K}\rceil$, $k = i - (b-1)K$, $n = j - BK - (k - 1)N_R$, and $m = l - BK - KN_R - (k - 1)M$.

Since most solvers can work only on \ac{QUBO} problems directly, we first convert the \ac{HUBO} problem into a constrained quadrature binary optimization problem. This conversion can be done by quadratization using the Rosenberg's procedure \cite{Anthony17}. Let $f(\xi_1, \ldots, \xi_{K(B+N_R+M)})$ be the function to be minimized with third-order monomials. The quadratization can be done iteratively as
\begin{enumerate}
        \item Select two variables $\xi_i$ and $\xi_j$ such that their product is the most common among the monomials of degree at least 3 in $f$.
        \item Let $h_{ij}$ be the function obtained upon replacing each occurrence of $\xi_i\xi_j$ by a new variable $\rho_{ij}$ in $f$.
        \item Let $g_{ij} = h_{ij} + P(\xi_i\xi_j - 2\xi_i\rho_{ij} - 2\xi_j\rho_{ij} + 3\rho_{ij})$, where $P$ is large enough positive number.
        \item Let $f = g_{ij}$. If the order of $f$ is higher than two, go to step 2.
\end{enumerate}

In the \ac{SU-MIMO} case to simplify the constraint mapping, we restrict the Rosenberg's procedure such that only $\xi_i$ and $\xi_j$, where $i = 1,\ldots,BK$ and $j = BK + 1,\ldots,BK + KN_R$, are selected for variable pairs. This can be interpreted such that auxiliary variable $\rho_{ij}$ corresponds to scheduling \ac{RB} $b = \lceil{i/K}\rceil$ to \ac{UE} $k = i - (b-1)K$ with $n = j - BK - (k - 1)N_R$ data streams. The number of auxiliary variables after the quadratization procedure is finished is $BKN_R$ and the total number of variables is $K(B + N_R + M + BN_R)$. In practice, the \ac{SNR} distribution is often such that no data can be transmitted with the highest \ac{MCS} indices, even with a single stream. Using (\ref{eq_Mmax}), we define the highest feasible \ac{MCS} index for \ac{UE} $k$ as $M_k = m_{\text{max},k}$ when $n = 1$. To reduce the number of \ac{QUBO} variables, only the \ac{MCS} indices $\{1,\ldots,M_k\}$ can be considered for \ac{UE} $k$. In this case, the number of \ac{QUBO} variables is $K(B + N_R + BN_R) + \sum\limits_{k=1}^K M_k$.

The \ac{QUBO} variable vector after quadratization is given as
\begin{equation}
    \begin{split}
        \hat{\mathbf{\xi}}^T = [&\mathbf{\xi}^T \quad \rho_{1,BK+1} \cdots \rho_{1,BK+N_R} \cdots\\ 
        &\rho_{K,BK+(K-1)N_R+1} \cdots \rho_{K,BK+KN_R}\\
        &\rho_{K+1,BK+1} \cdots \rho_{K+1,BK+N_R} \cdots\\
        &\rho_{2K,BK+(K-1)N_R+1} \cdots \rho_{2K,BK+KN_R} \cdots\\
        &\rho_{(B-1)K+1,BK+1} \cdots \rho_{(B-1)K+1,BK+N_R} \cdots\\
        &\rho_{BK,BK+(K-1)N_R+1} \cdots \rho_{BK,BK+KN_R}]^T
    \end{split}
\end{equation}
where the auxiliary variables are organized \ac{RB}-by-\ac{RB}, i.e. the first $KN_R$ elements after $\mathbf{\xi}$ correspond to \ac{RB} 1 and the last $KN_R$ elements in $\hat{\mathbf{\xi}}$ correspond to \ac{RB} $B$. The monomial coefficients are collected to matrix $\hat{\mathbf{Q}}$ after quadratization.

The linear constraints to the \ac{SU-MIMO} binary optimization problem are given in (\ref{constraint1_SUMIMO}), (\ref{Constraint3_SUMIMO}), and (\ref{Constraint2_SUMIMO}). In addition, there is an additional constraint for the auxiliary variables ensuring that only one user-number of streams pair is selected per \ac{RB}:
\begin{equation}
    \sum_{i=J+(b-1)KN_R+1}^{J+bKN_R} \hat{\xi}_i \leq 1, \quad \forall b = 1,\ldots,B
\end{equation}
where $J = K(B+N_R+M)$. Using the same transform as in Section \ref{sec_muMiso} for converting linear constraints to quadratic penalties, the \ac{QUBO} problem becomes
\begin{equation}
    \label{eq_suMimoQubo}
    \begin{split}
        \min_{\mathbf{\xi}} \hat{\mathbf{\xi}}^T \tilde{\mathbf{Q}} \hat{\mathbf{\xi}} &= \hat{\mathbf{\xi}}^T \hat{\mathbf{Q}} \hat{\mathbf{\xi}} + \sum_{i = (b-1)K+1}^{bK-1} \sum_{j = i + 1}^{bK} \lambda_1 \hat{\xi}_i \hat{\xi}_j + \\
        &\sum_{l = BK + (k-1)N_R + 1}^{BK + kN_R - 1} \sum_{o = l + 1}^{BK + kN_R} \lambda_2 \hat{\xi}_l \hat{\xi}_o + \\
        &\sum_{p=K(B+N_R)+(k-1)M+1}^{K(B+N_R)+kM-1} \sum_{q=p+1}^{K(B+N_R)+kM} \lambda_3 \hat{\xi}_p \hat{\xi}_q + \\
        &\sum_{s=J+(b-1)KN_R+1}^{J+bKN_R-1} \sum_{t=s+1}^{J+bKN_R} \lambda_4 \hat{\xi}_s \hat{\xi}_t.
    \end{split}
\end{equation}

\section{Numerical results}

In order to verify the correctness of the \ac{QUBO} formulations from Section \ref{sec_qubo} and to compare the complexity-performance trade-off of scheduling algorithms from Section \ref{sec_alg}, we simulate the system with the assumptions given in Table \ref{table_assumptions}. As already mentioned in Section \ref{sec_system}, constant fading is assumed within an \ac{RB}. We denote the channel matrix for user $k$ at \ac{RB} $b$ and \ac{TTI} $t$ as $\mathbf{H}_{k,b}(t)$.

\begin{table}[htbp]
\caption{Assumptions and parameter values used in simulations}
\begin{center}
\begin{tabular}{|c|c|}
\hline
MCS table& Table 5.1.3.1-2 \cite{TS38214}, $M=28$\\
\hline
\ac{SNR} thresholds&$\theta_m = 2^{\eta_m} - 1$ where $\eta_m$ is the spectral\\
&efficiency from Table 5.1.3.1-2 \cite{TS38214} for \ac{MCS} $m$\\
\hline
Transmitted \ac{SNR}&$P_T/N_0 = 10$\\
\hline
Fading assumptions&$\mathbf{H}_{k,b}(t) = (\sqrt{2}L_k)^{-1}\hat{\mathbf{H}}_{k,b}(t)$ where\\
&$L_k \sim U(1,3.981)$, $\hat{H}_{i,j,k,b}(t) \sim \mathcal{CN}(0,1)$,\\
&$\hat{\mathbf{H}}_{k,b}(t) \perp\!\!\!\perp \hat{\mathbf{H}}_{k+1,b}(t)$, $\hat{\mathbf{H}}_{k,b}(t) \perp\!\!\!\perp \hat{\mathbf{H}}_{k,b+1}(t)$,\\
&$\hat{\mathbf{H}}_{k,b}(t) \perp\!\!\!\perp \hat{\mathbf{H}}_{k,b}(t+1)$,\\
&uncorrelated spatial fading\\
\hline
Number of \acp{TTI} & $T_c = 100$\\
\hline
Proportional&Scaling at \ac{TTI} $t$ $\tilde{R}_k = \tilde{R}_k(t-1) =$\\
fairness&$\frac{T_c-1}{T_c}\tilde{R}_k(t-2) + \frac{1}{T_c}R_k(t-1)$ where\\
&$R_k(t-1)$ is the number scheduled data bits for\\
&user $k$ at previous \ac{TTI}\\
\hline
Traffic assumptions & Per-user \ac{DL} buffers never emptied\\
 & All traffic with the same priority\\
\hline
\end{tabular}
\label{table_assumptions}
\end{center}
\end{table}

\subsection{Empirical evaluation of the correctness of the QUBO formulations}

A straightforward way to verify the correctness of the \ac{QUBO} formulations from Section \ref{sec_qubo} would be to compare their solutions to the results from the corresponding optimal brute force algorithms of Section \ref{sec_alg}. However, this approach is infeasible in practice as the number of \ac{QUBO} variables grows beyond the capabilities of any optimal \ac{QUBO} solver. In addition, it soon becomes infeasible to obtain the optimal solutions from the brute force algorithm due to long execution time when $K > 4$. For cases where $K \leq 4$, it is possible to evaluate the correctness of the \ac{QUBO} matrices with the two tests: 1) When the brute force solution is converted to binary vectors $\hat{\mathbf{\xi}}$ and multiplied with the \ac{QUBO} matrix as $\hat{\mathbf{\xi}}^T \tilde{\mathbf{Q}} \hat{\mathbf{\xi}}$, we should get exactly the same result as the value of the objective function from the brute force search, and 2) Sub-optimal \ac{QUBO} solvers should never get better results than the brute force search.

When going through all the parameter pairs $B = \{10, \ldots, 273\}$ and $K = \{2, 3, 4\}$, the multiplication in the first test resulted in the same value as the optimal objective function value for all cases. For the second test, the formulated \ac{QUBO} is solved using the simulated annealing and simulated quantum annealing algorithms from D-Wave Ocean SDK \cite{DWave26}.

\begin{figure}[htbp]
\centerline{\includegraphics[width=\columnwidth]{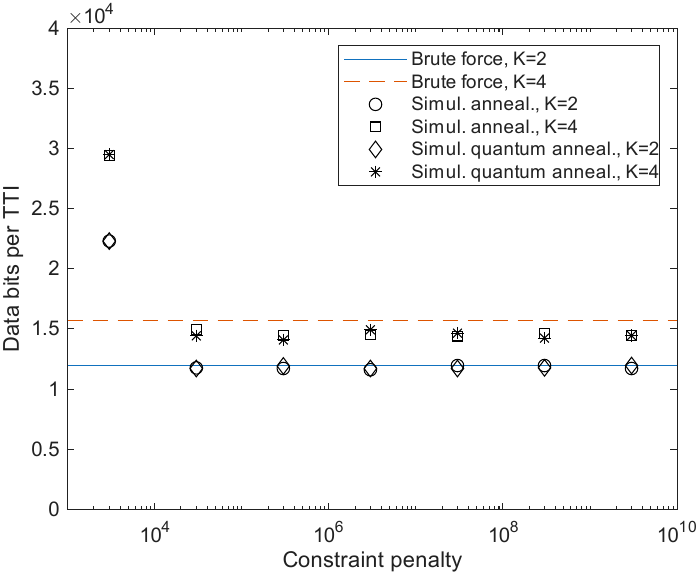}}
\caption{The number of scheduled data bits per \ac{TTI} as a function of $\lambda_i$, \ac{SU-MIMO}, $B = 20$, $N_T = N_R = 4$.}
\label{fig_lambdaEffect}
\end{figure}

\begin{figure}[htbp]
\centerline{\includegraphics[width=\columnwidth]{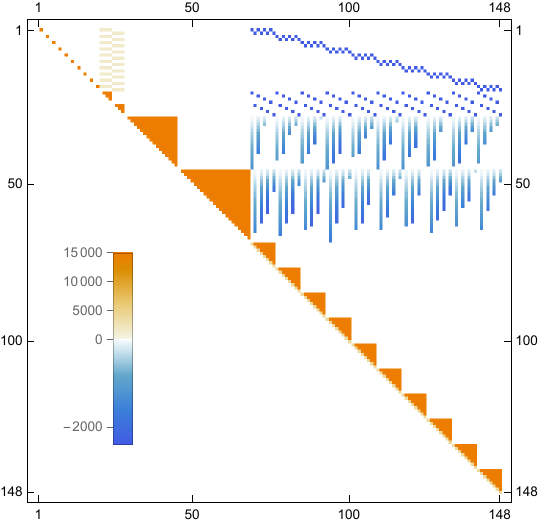}}
\caption{Heatmap of an example upper triangular \ac{QUBO} matrix, \ac{SU-MIMO}, $B = 10$, $K = 2$, $N_R = 4$.}
\label{fig_quboHeatMap}
\end{figure}

The parameters for solving the \ac{QUBO} problems are given in Table \ref{table_solver}. Different parameter values were tested until a set that worked fairly well with simulated annealing algorithms was found. The quadratization penalty $P$ was set to a slightly larger value than the maximum number of data bits for an \ac{RB} with a single data stream. The constraint penalty $\lambda_i$ was set to scale with $B$ like the optimal solution as recommended in \cite{Glover19}. The simulated annealing algorithms were not very sensitive to the selection of the constraint penalty as long as the penalty was large enough. This is illustrated in Fig. \ref {fig_lambdaEffect} where the number of scheduled data bits per \ac{TTI} are shown as a function of $\lambda_i$ (the same $\lambda_i$ is used for all constraints) for \ac{SU-MIMO} when $N_T = N_R = 4$ and $B=20$. It can be seen that the constraints are violated, which is visible as solutions exceeding the optimum value, when too small penalty values are used. An example \ac{QUBO} matrix presented as a heat map is shown in Fig. \ref{fig_quboHeatMap}. The dark orange elements correspond to the penalties from the constraints, while the other elements result from the quadratization procedure. 

\begin{table}[htbp]
\caption{QUBO solver parameters}
\begin{center}
\setlength{\tabcolsep}{3pt}
\begin{tabular}{|c|c|}
\hline
Penalty for constraints&$\lambda_i = 1000B$, $\forall i$, \ac{MU-MISO}\\
 &$\lambda_i = 1500B$, $\forall i$, \ac{SU-MIMO}\\
\hline
Quadratization penalty (\ac{SU-MIMO})&$P = 1200$\\
\hline
Temperature range and schedule&Default, set by Ocean SDK\\
\hline
Number of reads&500\\
\hline
Number of sweeps&$20000$\\
\hline
Number of sweeps per temperature&2\\
\hline
\end{tabular}
\label{table_solver}
\end{center}
\end{table}

\begin{figure}[htbp]
\centerline{\includegraphics[width=\columnwidth]{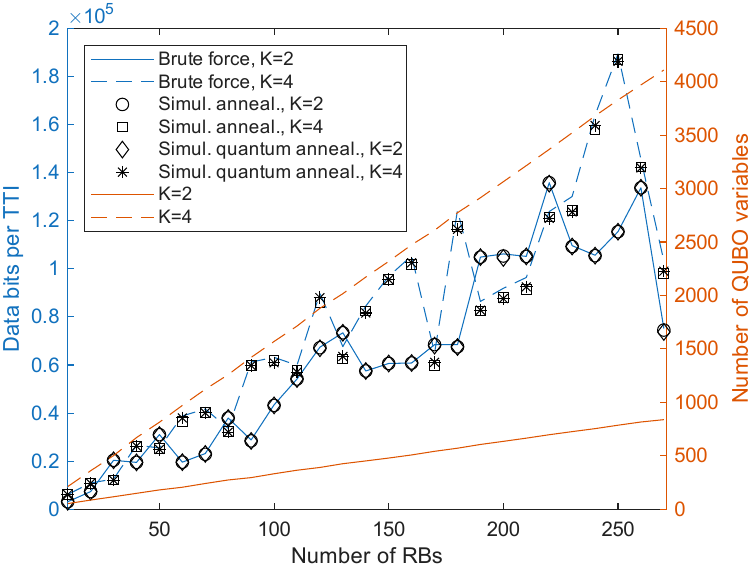}}
\caption{The number of scheduled data bits per \ac{TTI} and the number of \ac{QUBO} variables as a function of number of \acp{RB}, \ac{MU-MISO}, $N_T = 4$.}
\label{fig_quboMuMiso}
\end{figure}

\begin{figure}[htbp]
\centerline{\includegraphics[width=\columnwidth]{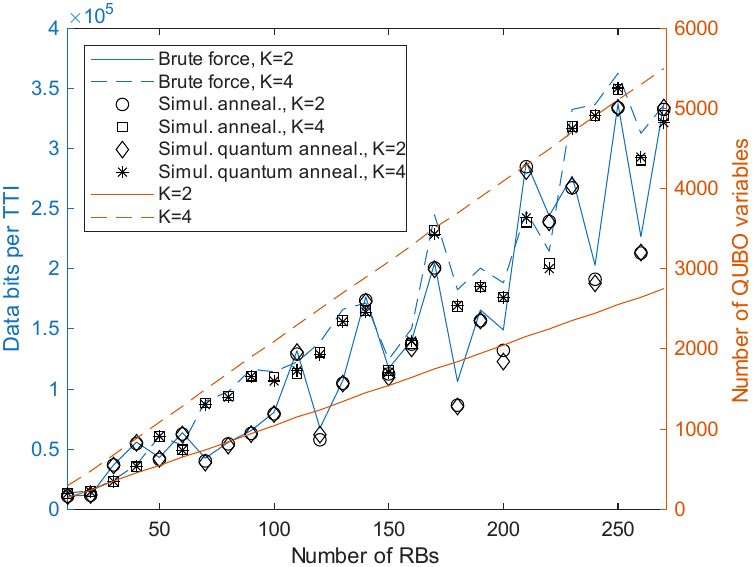}}
\caption{The number of scheduled data bits per \ac{TTI} and the number of \ac{QUBO} variables as a function of number of \acp{RB}, \ac{SU-MIMO}, $N_T = N_R = 4$.}
\label{fig_quboSuMimo}
\end{figure}

The results of the \ac{QUBO} formulation evaluation are shown Figs. \ref{fig_quboMuMiso} and \ref{fig_quboSuMimo} where solutions for a single \ac{TTI} are shown for \ac{MU-MISO} and \ac{SU-MIMO}, respectively. The number of base station transmitters is fixed to $N_T = 4$ for both cases, and the number of receiving antennas is fixed to $N_R = 4$ for the \ac{SU-MIMO} case. It can be seen that simulated annealing is able to achieve close-to-optimal solutions. In addition, the solutions from the simulated annealing should never exceed those from the brute force search, which is the case in Figs. \ref{fig_quboMuMiso} and \ref{fig_quboSuMimo}. This, together with the results from the first test, gives us confidence that the \ac{QUBO} matrices in (\ref{eq_muMisoQubo}) and (\ref{eq_suMimoQubo}) are correctly formulated. The number of \ac{QUBO} variables as a function of the number of \acp{RB} is also shown as red lines. This illustrates that the number of \ac{QUBO} variables increase linearly with the increasing bandwidth (number of \acp{RB}) for both \ac{MU-MISO} and \ac{SU-MIMO}.

\begin{figure}[htbp]
\centerline{\includegraphics[width=\columnwidth]{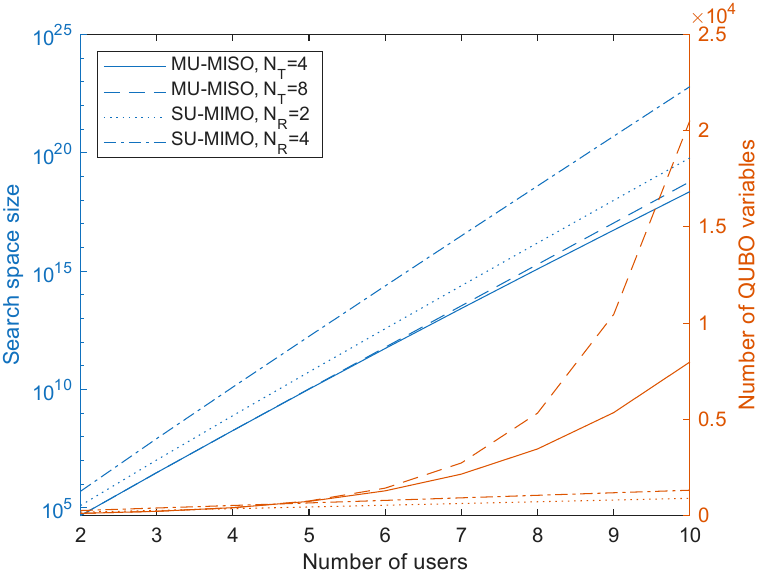}}
\caption{The search space size and the number of \ac{QUBO} variables as a function of number of users, $B = 20$.}
\label{fig_searchSize}
\end{figure}

The scaling of the problem search space size and the number of \ac{QUBO} variables are shown in Fig. \ref{fig_searchSize} as a function of the number of users when $B = 20$. This illustrates the problem complexity increases exponentially as $K$ grows for both \ac{MU-MISO} and \ac{SU-MIMO} cases. The \ac{QUBO} formulation of the \ac{MU-MISO} scheduling problem is practical for only a small number of users as the number of required \ac{QUBO} variables grows polynomially with $K^{N_T}$. On the other hand, the number of \ac{QUBO} variables grows only linearly with the \ac{SU-MIMO} \ac{QUBO} formulation. 

Although the \ac{SU-MIMO} \ac{QUBO} formulation looks promising in terms of the number of \ac{QUBO} variables, the coupling between the variables can be problematic for quantum implementation. To study this further, we formulated the \ac{SU-MIMO} \ac{QUBO} matrices as logical graphs and attempted to embed them into the topologies of the current state-of-the-art D-Wave quantum annealers. The results indicate that even though the embedding can be done when both the number of \acp{UE} and \acp{RB} are low, some of the logical \ac{QUBO} variables have to be represented by long chains of physical qubits. This makes the practical implementation of the \ac{SU-MIMO} scheduling infeasible with the current state-of-the-art quantum annealers, see Appendix \ref{app_embedding} for details. However, the qubit connectivity has been improving with each new topology. If the same trend continues, embedding to the future quantum annealers should be easier with shorter chain lengths.

\subsection{Complexity and performance of the scheduling algorithms}

Based on the scalability analysis, the \ac{SU-MIMO} \ac{QUBO} formulation looks promising for further study because the number of \ac{QUBO} variables grows only linearly while the problem search space grows exponentially. However, even if we had a quantum solver capable of solving the \ac{SU-MIMO} problem optimally with a high probability, a relevant question is whether solving the \ac{SU-MIMO} problem is worth the effort. In other words, how close to optimum we can get with the low-complexity sub-optimal algorithms.

To get a better understanding of this question, we analyze the sub-optimal algorithms presented in Section \ref{sec_alg_sumimo} in terms of their complexity and performance. The considered algorithms are the exhaustive brute force search, the brute force search with Constraint (\ref{Constraint3_SUMIMO}) relaxed and \ac{MCS} indices decided after scheduling, and the greedy search. The algorithms are labeled as 'Brute force', 'No MCS', and 'Greedy' in the figures of this section, respectively. 

\begin{figure}[htbp]
    \centering
    \subfloat[]{\includegraphics[width=0.497\columnwidth]{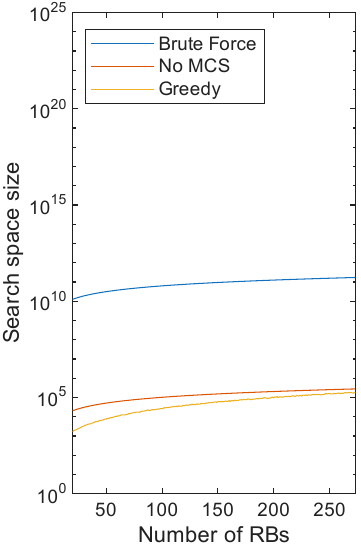}
    \label{fig_searchSizeAlgorithms}}
    \hfil
    \subfloat[]{\includegraphics[width=0.474\columnwidth]{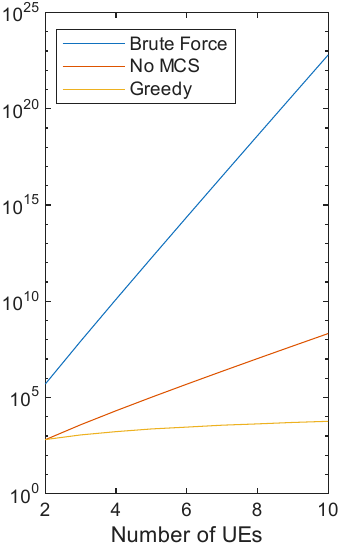}
    \label{fig_searchSizeAlgorithms_numUe}}
    \caption{(a) Search space size for the \ac{SU-MIMO} algorithms as a function of $B$, $K=4$, $N_R = 4$. (b) Search space size for the \ac{SU-MIMO} algorithms as a function of $K$, $B=20$, $N_R = 4$.}
    \label{SplittedFigure}
\end{figure}

The algorithm complexity is presented as the search space size assuming that all $M$ \ac{MCS} indices are included in the search for all \acp{UE}. The search space size is shown as a function of number of \acp{RB} and number of \acp{UE} in Fig. \ref{SplittedFigure}, (\ref{fig_searchSizeAlgorithms}) and (\ref{fig_searchSizeAlgorithms_numUe}), respectively. As explained in Section \ref{sec_alg_sumimo}, the search space size is random for the greedy algorithm, and thus each point for the greedy algorithm represents an average over 100 time slots; the variability of the average is not visible on the logarithmic scale. The search space size of the greedy algorithm grows only linearly with $K$, which makes it a potential option for real-time scheduling at each time slot.

\begin{figure}[htbp]
\centerline{\includegraphics[width=\columnwidth]{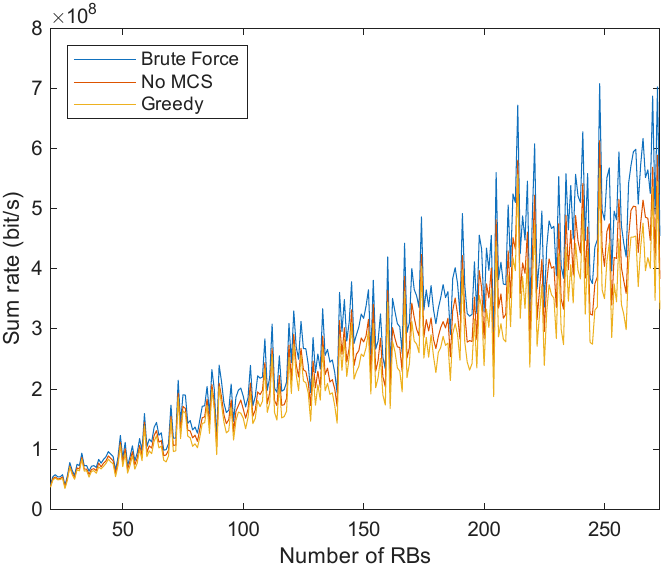}}
\caption{Sum rate of the \ac{SU-MIMO} algorithms as a function of $B$, $K = N_R = N_T = 4$.}
\label{fig_sumRate}
\end{figure}

The performance of the \ac{SU-MIMO} algorithms in terms of the sum rate over 100 time slots as a function of number of \acp{RB} is shown in Fig. \ref{fig_sumRate}. As expected, the brute force algorithm performs best while the greedy algorithm achieves the lowest sum rate. The sum rate curves do not increase monotonically because the path loss terms $L_k$ in the channel matrix (see Table \ref{table_assumptions}) are randomly drawn from the uniform distribution for each value of $B$ causing considerable variation in the 'goodness' of the \ac{UE} locations.

\begin{figure}[htbp]
\centerline{\includegraphics[width=\columnwidth]{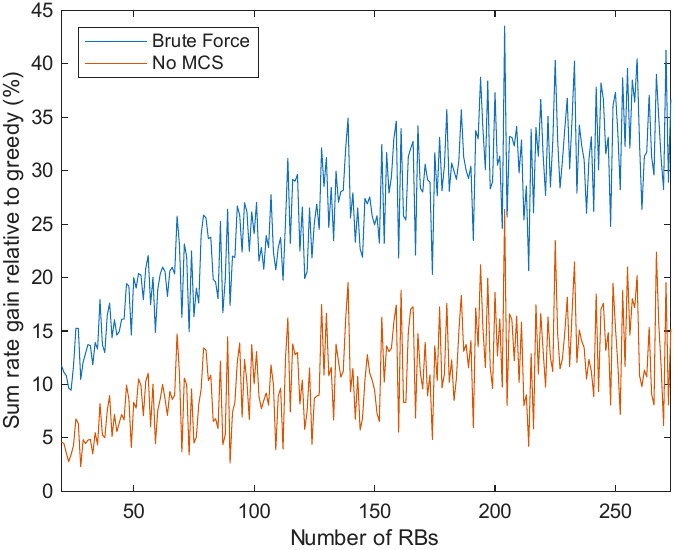}}
\caption{Sum rate gain relative to the greedy algorithm as a function of $B$, $K = N_R = N_T = 4$.}
\label{fig_averageGain}
\end{figure}

To make it easier to interpret the gain of using the optimal brute force algorithm, the sum rate gain relative to the greedy algorithm is shown in Fig. \ref{fig_averageGain}. The performance gain of using the optimal brute force algorithm compared to the greedy algorithm when  $K=N_R=N_T=4$ is 10-40\% with the growing trend with respect to the number of \acp{RB}.

\begin{figure}[htbp]
\centerline{\includegraphics[width=\columnwidth]{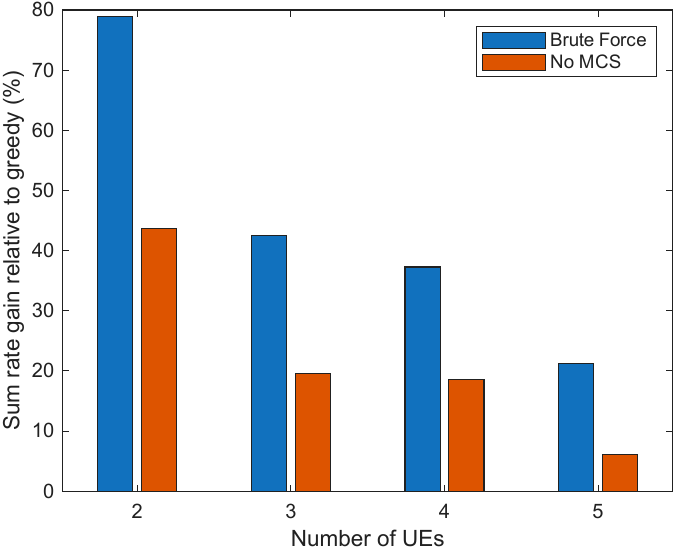}}
\caption{Sum rate gain relative to the greedy algorithm as a function of $K$, $B = 200$, $N_R = N_T = 4$.}
\label{fig_averageGain_K}
\end{figure}

When the number of \acp{UE} considered for scheduling increases, the performance gap between the greedy and brute force algorithms decreases. This is illustrated in Fig. \ref{fig_averageGain_K} where the sum rate gain relative to the greedy algorithm is shown for several values of $K$ when $B = 200$. We had to limit the number of \acp{UE} up to $K=5$ because the brute force algorithm run times grow very long with higher values of $K$. For example with $K=6$, $B=200$, $N_R=4$, the search space size is $2.4\cdot10^{15}$. Even if the gain from the optimal brute force algorithm is small for large $K$ in a single time slot, the cumulative effect of being able to serve more traffic and to keep the user-experienced quality of service high might still be notable in the long term.

The performance results in this section indicate that the gain from solving the \ac{SU-MIMO} scheduling optimally is largest when the number of scheduled \acp{UE} is small and the number of \acp{RB} is large. Assuming that we had a working quantum solver for the \ac{QUBO} problem available, it would make sense to pre-evaluate the limits for the pair of parameters $K$ and $B$ to decide when to use the \ac{QUBO}-based optimum solver and when a suboptimal low-complexity algorithm, such as the greedy search.

\section{Conclusion}

We have formulated the \ac{DL} \ac{MU-MISO} and \ac{SU-MIMO} scheduling problems in 5G base stations as \ac{QUBO} and analyzed their scalability in terms of the number of required \ac{QUBO} variables. In particular, the \ac{SU-MIMO} \ac{QUBO} formulation looks promising for further study as the number of \ac{QUBO} variables grows only linearly as a function of number of users while the problem search space size grows exponentially. This is important as there can be tens of users considered for scheduling in the network during the rush hours. 

In addition, we analyze the complexity-performance trade-off of suboptimal scheduling algorithms, which provides insight into when solving the problem optimally using quantum computers may be useful in practice. The results indicate that the gain from solving the \ac{SU-MIMO} \ac{DL} scheduling problem optimally is largest when the number of resource blocks is large and the number of users is small. In practice, either a \ac{QUBO}-based optimum solver or a suboptimal low-complexity algorithm could be selected based on system parameters.

Our initial embedding study shows that it is not yet feasible to embed the \ac{SU-MIMO} \ac{QUBO} graphs to the current state-of-the-art D-Wave quantum annealers when the number of resource blocks is 70 or larger. In addition, the chain lengths after embedding are so long that it would be very unlikely to get close-to-optimal solutions from the current annealers. However, our results can be used as a starting point for alternative and potentially more practical \ac{QUBO} formulations of the scheduling problem as well as for quantum implementations. A potential future work could to be to apply the \ac{QUBO} approach only on \ac{MU-MIMO} subproblems such as virtual user selection.

\appendices
\section{Embedding to D-Wave topologies}
\label{app_embedding}

To evaluate the practical feasibility of applying the proposed \ac{QUBO} formulations on current quantum annealers, we study the minor embedding of the \ac{QUBO} matrices derived in Section \ref{sec_qubo} onto D-Wave hardware topologies.
Embedding maps the logical \ac{QUBO} variables to the physical qubits of the given quantum processing unit topology. Because the practical qubit topologies are not fully connected, some of the logical variables must be represented as a chain of physical qubits. Longer chains are generally undesirable because they increase the likelihood of chain breaks reducing the quality of the solutions returned by a quantum annealer \cite{DWave26c}.

We focus on the \ac{SU-MIMO} case, which is the most interesting for practical implementation due to the linear scaling of the number of \ac{QUBO} variables. In addition for the rest of this section, we assume $N_T=N_R=4$. \ac{QUBO} matrices $\tilde{\mathbf{Q}}$ from (\ref{eq_suMimoQubo}) can be represented as logical \ac{QUBO} graphs $G_Q = (V_Q,E_Q)$ where each binary variable corresponds to a vertex and an edge is added between variables $i$ and $j$ whenever $\tilde{q}_{ij} \neq 0, i \neq j$. Before actual embedding, its difficulty can be estimated by evaluating the \ac{QUBO} graph degree and density. The summary of the results from the \ac{QUBO} graph analysis is collected to Table \ref{table_graph}. The values are the averages from 100 individual channel realizations per each $\{K,B\}$ pair. This approach was selected because, as discussed in Section \ref{sec_suMimoQubo}, there is some variation in the number of logical \ac{QUBO} variables due to the variation in the number of feasible \ac{MCS} indices. It can be seen from Table \ref{table_graph} that the the graph density decreases as both $K$ and $B$ increase. In addition, the average degree increases only slightly with the increasing problem size. This indicates that the \ac{QUBO} graphs remain relatively sparse as the problem scales, which is favorable for minor embedding onto quantum annealing hardware. However, the maximum degree of the graphs is high. This necessitates the use of long chains of physical qubits with real-life quantum annealer topologies.

\begin{table}[htbp]
\caption{QUBO graph properties}
\begin{center}
\setlength{\tabcolsep}{3pt}
\begin{tabular}{|c|c|c|c|c|c|c|}
\hline
$K$&$B$&$|V_Q|$&$|E_Q|$&Graph density&Avg. degree&Max. degree\\
\hline
2&20&246&3145&0.104&25.5&95.2\\
\hline
2&30&346&4474&0.0749&25.8&132\\
\hline
2&40&447&5991&0.0599&26.8&171\\
\hline
2&50&547&7435&0.0496&27.1&209\\
\hline
2&60&647&8665&0.0415&26.8&245\\
\hline
2&70&747&9969&0.0358&26.7&280\\
\hline
3&20&369&5226&0.0769&28.3&97.3\\
\hline
3&30&519&7505&0.0558&28.9&135\\
\hline
3&40&669&9866&0.0440&29.4&173\\
\hline
4&20&492&7661&0.0634&31.1&99.2\\
\hline
4&30&692&11079&0.0463&32.0&137\\
\hline
5&20&612&10186&0.0544&33.3&98.1\\
\hline
\end{tabular}
\label{table_graph}
\end{center}
\end{table}

The target topologies for embedding are selected as Pegasus $P_{16}$ and Zephyr $Z_{12}$ that correspond to D-Wave Advantage and Advantage2 quantum annealers, respectively \cite{DWave26b}. A logical \ac{QUBO} graph is randomly selected from the 100 generated graphs for each $\{K,B\}$ pair. The logical \ac{QUBO} graphs are embedded onto the target topologies using the find\_embedding function of the Ocean SDK minorminer library. The find\_embedding function is an implementation of the heuristic algorithm for finding graph minors presented in \cite{Cai14}. Because the minor embedding algorithm is heuristic, we repeat the embedding 10 times for each \ac{QUBO} graph and select the best embedding with the lowest qubit count. For each successful embedding, the number of needed physical qubits, the median chain length, and the 75th and 95th percentile of the chain length are collected to Tables \ref{table_embedding_Pegasus} and \ref{table_embedding_Zephyr} for the Pegasus $P_{16}$ and Zephyr $Z_{12}$ topologies, respectively. In general, the Zephyr $Z_{12}$ topology with the improved qubit connectivity achieves shorter chain lengths than the Pegasus $P_{16}$ topology. Unfortunately, for both topologies and for all the tried $\{K,B\}$ pairs there are many long chains. This either requires the use of high chain strengths or increases the probability of chain breaks, both of which significantly decrease the probability of achieving close-to-optimal solutions \cite{DWave26c}. When $K$ or $B$ increases, the embedding can no longer be done because there are not enough physical qubits available. This happens e.g. when $K=2$ and $B \geq 70$.

\begin{table}[htbp]
\caption{Embedding results for the Pegasus $P_{16}$ topology}
\begin{center}
\setlength{\tabcolsep}{3pt}
\begin{tabular}{|c|c|c|c|c|c|c|c|}
\hline
$K$&$B$&$|V_Q|$&$|E_Q|$&Qubits&Chain length&Chain length&Chain length\\
&&&&&median&P75&P95\\
\hline
2&20&247&3224&1601&5&8&15\\
\hline
2&30&336&3417&2328&5&7&25\\
\hline
2&40&449&6367&3885&6&9&30\\
\hline
2&50&543&6495&4313&5&7&37\\
\hline
2&60&642&7622&4819&4&6&40\\
\hline
2&70&750&10536&N/A&N/A&N/A&N/A\\
\hline
3&20&367&5055&3745&9&14&22\\
\hline
3&30&506&6121&4668&5&8&42\\
\hline
3&40&667&9410&N/A&N/A&N/A&N/A\\
\hline
4&20&497&8027&N/A&N/A&N/A&N/A\\
\hline
\end{tabular}
\label{table_embedding_Pegasus}
\end{center}
\end{table}

\begin{table}[htbp]
\caption{Embedding results for the Zephyr $Z_{12}$ topology}
\begin{center}
\setlength{\tabcolsep}{3pt}
\begin{tabular}{|c|c|c|c|c|c|c|c|}
\hline
$K$&$B$&$|V_Q|$&$|E_Q|$&Qubits&Chain length&Chain length&Chain length\\
&&&&&median&P75&P95\\
\hline
2&20&247&3224&1345&5&7&12\\
\hline
2&30&336&3417&1782&4&5&20\\
\hline
2&40&449&6367&3136&4&6&33\\
\hline
2&50&543&6495&3438&4&5&34\\
\hline
2&60&642&7622&3938&4&5&29\\
\hline
2&70&750&10536&N/A&N/A&N/A&N/A\\
\hline
3&20&367&5055&3236&7&11&24\\
\hline
3&30&506&6121&3792&5&7&34\\
\hline
3&40&667&9410&N/A&N/A&N/A&N/A\\
\hline
4&20&497&8027&N/A&N/A&N/A&N/A\\
\hline
\end{tabular}
\label{table_embedding_Zephyr}
\end{center}
\end{table}

\section*{Acknowledgment}

The authors want to thank Hannu Reittu for his early assistance with the QUBO formulations, Kari Seppänen for CPLEX testing, Prof. Animesh Yadav for many  fruitful discussions on the research topic, and Costantino Carugno and Andrea Marchesin for help and feedback in quantum annealing questions.

\bibliographystyle{IEEEtran}
\bibliography{IEEEabrv,cohqcaSchedulingBibliography}

\end{document}